\documentclass{aa}  

\usepackage{bm}
\usepackage{color}
\usepackage{float}
\usepackage{graphicx,amsmath}
\usepackage{txfonts}
\usepackage{natbib}
\def\wig#1{\mathrel{\hbox{\hbox to 0pt{%
          \lower.6ex\hbox{$\sim$}\hss}\raise.4ex\hbox{$#1$}}}}

\def\mearth{{\rm M_\oplus}}

\def\mdotf{\dot{M}_{\rm peb}}
\def\mdots{\dot{M}_*}
\def\mf4{\dot{M}_{\rm F4}}

\def\au{{\rm au}}
\def\hg{h_{\rm g}}
\def\hp{h_{\rm p}}
\def\vr{{\rm v}_r}  
\def\st{\tau_{\rm s}}
\def\stpeb{\tau_{\rm s,peb}}
\def\sigp{\Sigma_{\rm p}}
\def\sigg{\Sigma_{\rm g}}
\def\vk{{\rm v}_{\rm K}}

\usepackage[normalem]{ulem}

\graphicspath{{figs/}{./}}
\DeclareGraphicsExtensions{.pdf,.png,.jpg}

\begin{document}

   \title{Formation of dust-rich planetesimals from sublimated pebbles inside of the snow line}

   \author{S. Ida
          \inst{1}
          \and
          T. Guillot
          \inst{2}
           }

   \institute{Earth-Life Science Institute, Tokyo Institute of Technology, Meguro-ku, Tokyo 152-8550, Japan\\
              \email{ida@elsi.jp}
         \and
             Laboratoire J.-L.\ Lagrange, Universit\'e C\^ote d'Azur, Observatoire de la
  C\^ote d'Azur, CNRS, F-06304 Nice, France\\
           }

   \date{DRAFT:  \today}


  \abstract
{For up to a few millions of years, pebbles must provide a quasi-steady inflow of solids from the outer parts of protoplanetary disks to their inner regions. }
   {We wish to understand how a significant fraction of the pebbles grows into planetesimals instead of being lost to the host star. }
   {We examined analytically how the inward flow of pebbles is affected by the snow line and under which conditions 
    dust-rich (rocky) planetesimals form. When calculating the inward drift of solids that is due to gas drag, we included the back-reaction of the gas to the motion of the solids. }
   {We show that in low-viscosity protoplanetary disks (with a monotonous surface density similar to that of the minimum-mass solar nebula), the flow of pebbles does not usually reach the required surface density to form planetesimals by streaming instability. We show, however, that if the pebble-to-gas-mass flux exceeds a critical value, no steady solution can be found for the solid-to-gas ratio. This is particularly important for low-viscosity disks ($\alpha<  10^{-3}$) where we show that inside of the snow line, silicate-dust grains ejected from sublimating pebbles can accumulate, eventually leading to the formation of dust-rich planetesimals directly by gravitational instability. 
}
   {This formation of dust-rich planetesimals may occur for extended periods of time, while the snow line sweeps from several au to inside of 1 au. The rock-to-ice ratio may thus be globally significantly higher in planetesimals and planets than in the central star. 
}

     
     \titlerunning{Dust-rich planetesimals from sublimated pebbles}

   \maketitle
%

\section{Introduction}

Determining the fate of solids in protoplanetary disks is key for
understanding the birth and growth of planets and planetary systems.  
While small grains are coupled to the disk gas,
large particles drift inward as a consequence of angular momentum loss by aerodynamical gas drag.
For meter sizes (assuming compact grains), the inward drift velocity is $\sim 10^{-2}$ \au/yr
 \citep[e.g.,][]{Weiden80,Nakagawa81}. 
For small dust grains, growth through pairwise collisions is faster than drift
so that they grow {\it \textup{in situ}} until they reach $1$ to $100$
cm, at which point drift starts to dominate 
\citep[e.g.,][]{Okuzumi12,LJ14b}. 
These so-called pebbles then drift rapidly with limited growth, implying
that without a mechanism to suppress the drift, they would be lost
to the central star. 

Planetesimals would form directly by
gravitational instability (GI) in the dust disk if it is sufficiently
thin and dense \citep[e.g.,][]{GW73}.  However, even in low-turbulence
disks, the Kelvin-Helmholtz (KH) instability generated by the vertical
shear between the dust subdisk and the gas prevents the development
of a disk that is thin enough \citep[e.g.,][]{Weidenschilling95,Sekiya98}. 
\citet{YoudinShu02} proposed that migrating dust (or pebbles) would pile up
in the inner disk to become gravitationally unstable
\cite[see also][]{Laibe12}, but they
neglected grain growth, which was then shown to prevent this pile-up
\cite[see][]{Krijt16}. 

Another possibility to form planetesimals is to invoke streaming
instabilities (SI) in the drifting pebble flow: When their
density is high enough, clumps can form, and because they undergo
relatively less gas drag, they accrete individual pebbles to rapidly form planetesimals of 
100 to 1000\,km \citep{Youdin05,
  Johansen07}. However, this mechanism requires a high solid-to-gas
ratio ($Z$) and has also been shown to be difficult to achieve in realistic
disks \citep{Krijt16}.

With analytical calculations, we examine here the formation of planetesimals from drifting
pebbles in smooth disks (i.e., without pressure bumps, gaps, or
vortexes)  through these two mechanisms. 
We highlight the importance of sublimation across the snow line. 
Instead of examining the consequence of ice deposition beyond the snow line in turbulent disks \citep[e.g.,][]{Stevenson_Lunine,Ros_Johansen,Armitage+2016}, following \cite{saito_sirono11}, we  
concentrate on the region inside the snow line where dust grains ejected from sublimated pebbles are present.

\section{Pebble-to-gas surface density ratio}

We consider a protoplanetary disk characterized by a steady gas accretion rate $\mdots$ in which the solids are in the form of pebbles migrating inward at a mass flux $\mdotf$. 
The surface density of the migrating pebbles $\sigp$
and the disk gas $\sigg$ are given by
\begin{equation}
\begin{array}{ll}
\sigp & 
{\displaystyle = 
\mdotf/2\pi r \vr},\\
 \sigg & 
 {\displaystyle \simeq 
\dot{M}_*/3\pi \nu\simeq \dot{M}_*/3\pi \alpha \hg^2 \Omega_{\rm K}},
\label{eq:Sigma}
\end{array}
\end{equation}
where $\nu$ is the turbulent viscosity of the gas disk, $\hg$ the gas scale height,  
$\Omega_{\rm K}$ the Keplerian frequency, and $\vr$ is
the pebble migration speed.
We use the $\alpha$-prescription, that is, $\nu \simeq \alpha \hg^2 \Omega_{\rm K}$. 

The inward radial drift speed of solids was calculated in the limit of a static disk by \cite{Nakagawa86} and in the limit of a low solid-to-gas ratio by \cite{Guillot14}. Combining the two yields 
\begin{equation}
\vr = -\Lambda^2 \frac{2\st}{1+\Lambda^2 \st^2}\eta \vk
+ \Lambda \frac{1}{1+\Lambda^2 \st^2}u_\nu,
\label{eq:vr}
\end{equation}
where the back-reaction of the gas to the motion of solids has been included through $\Lambda \equiv \rho_{\rm g}/(\rho_{\rm g} + \rho_{\rm p})$, and $\rho_{\rm g}$ and $\rho_{\rm p}$ are the mid-plane densities of gas and solids, respectively. 
We included the $\Lambda$-dependence of $u_\nu$ as well for later purposes.
In Eq.~\eqref{eq:vr}, the size of the solids is defined through their Stokes number $\st$ , which is the ratio of their stopping time due to gas drag ($t_{\rm stop}$)
to the Kepler frequency as
\begin{equation}
\st = t_{\rm stop}\Omega_{\rm K},
\end{equation}
$u_\nu$ is the radial velocity of the accreting disk gas, which in the inner regions of a vertically uniform disk may be approximated by 
\begin{equation}
u_\nu \sim - 3\nu/2r \sim 
- 3 \alpha \hg^2 \Omega_{\rm K}/2r \sim - (3/2) \alpha (\hg/r)^2 \vk,
\label{eq:ur}
\end{equation}
and $\eta$ $(\ll 1)$ is the deviation fraction of the gas orbital angular velocity ($\Omega$)
relative to the Keplerian angular velocity ($\Omega_{\rm K}$) that is due to the radial pressure gradient in the disk,
\begin{equation}
\eta = \frac{\Omega_{\rm K}-\Omega}{\Omega_{\rm K}}\simeq
\frac{1}{2} \left(\frac{\hg}{r}\right)^2 \left( - \frac{d \ln P}{d \ln r} \right).
\label{eq:eta0}
\end{equation}
From Eqs.~(\ref{eq:vr}), (\ref{eq:ur}), and (\ref{eq:eta0}),
\begin{equation}
\vr \simeq - \frac{1}{1+\Lambda^2\st^2} \left[ \Lambda^2 \st
\left( -\frac{d \ln P}{d \ln r} \right) + \Lambda \frac{3\alpha}{2} \right] \left(\frac{\hg}{r}\right)^2 \vk.
\label{eq:vr2}
\end{equation}
Using Eq.~(\ref{eq:Sigma}), we then obtain the solid-to-gas ratio as
\begin{equation}
\begin{array}{ll}
{\displaystyle Z = \frac{\sigp}{\Sigma_g}} & 
{\displaystyle \simeq 
\frac{3 \alpha}{2} \left(\frac{\hg}{r}\right)^2 
\frac{\vk}{\vr} 
\frac{\mdotf}{\dot{M}_*}} \\
 &
{\displaystyle \simeq 
\left(1+\Lambda^2 \st^2\right)
\left[ \frac{2 \st}{3 \alpha} \Lambda^2 
\left(- \frac{d \ln P}{d \ln r} \right) + \Lambda \right]^{-1}
\frac{\mdotf}{\dot{M}_*}} .
\end{array}
\label{eq:solid_to_gas}
\end{equation}

Now, the parameter $\Lambda$ may be estimated in the limit of a vertically isothermal disk as 
\begin{equation}
\Lambda^{-1} = 1 + \frac{\sigp/\hp}{\sigg/\hg} = 1 + Z\frac{\hg}{\hp}.
\label{eq:Lambda}
\end{equation}
This thus leads to the following second-order equation in $Z$:
\begin{equation}
(\xi\beta^2-\beta)Z^2-(A+1-2\xi \beta)Z+(1+\tau_s^2)\xi=0,\label{eq:Z-eq}
\end{equation}
where we have defined a few quantities,
\begin{equation}
A \equiv 
\frac{2\st}{3\alpha} \left(-\frac{d\ln P}{d\ln r}\right)\equiv a_0(\st/\alpha),
\end{equation}
and we adopt $a_0\approx 1.75$ as $a_0$ is estimated to be $\approx 1.70-1.85$ for radially smooth disks with both viscous heating and stellar irradiation \citep[see][]{Ida16}, $\beta$ is the ratio of the gas-to-dust pressure scale heights \citep{Dubrulle95,Youdin_Lithwick07},
\begin{equation}
\beta \equiv \hg/\hp\simeq (1+\st/\alpha)^{1/2},
\end{equation}
and $\xi$ is the ratio of the solid mass flux to the gas mass flux:
\begin{equation}
\xi \equiv \dot{M}_{\rm peb}/ \dot{M}_*.
\end{equation}
The solutions to Eq.~\eqref{eq:Z-eq} are\begin{equation}
Z=\frac{2\xi\beta - (A+1) \pm \sqrt{(A+1)^2-4\xi\beta (A+\xi\beta\st^2-\st^2)}}{2\beta(1- \xi\beta)}.
\label{eq:Z}
\end{equation}

\begin{figure}[htb]
\includegraphics[width=\hsize,angle=0]{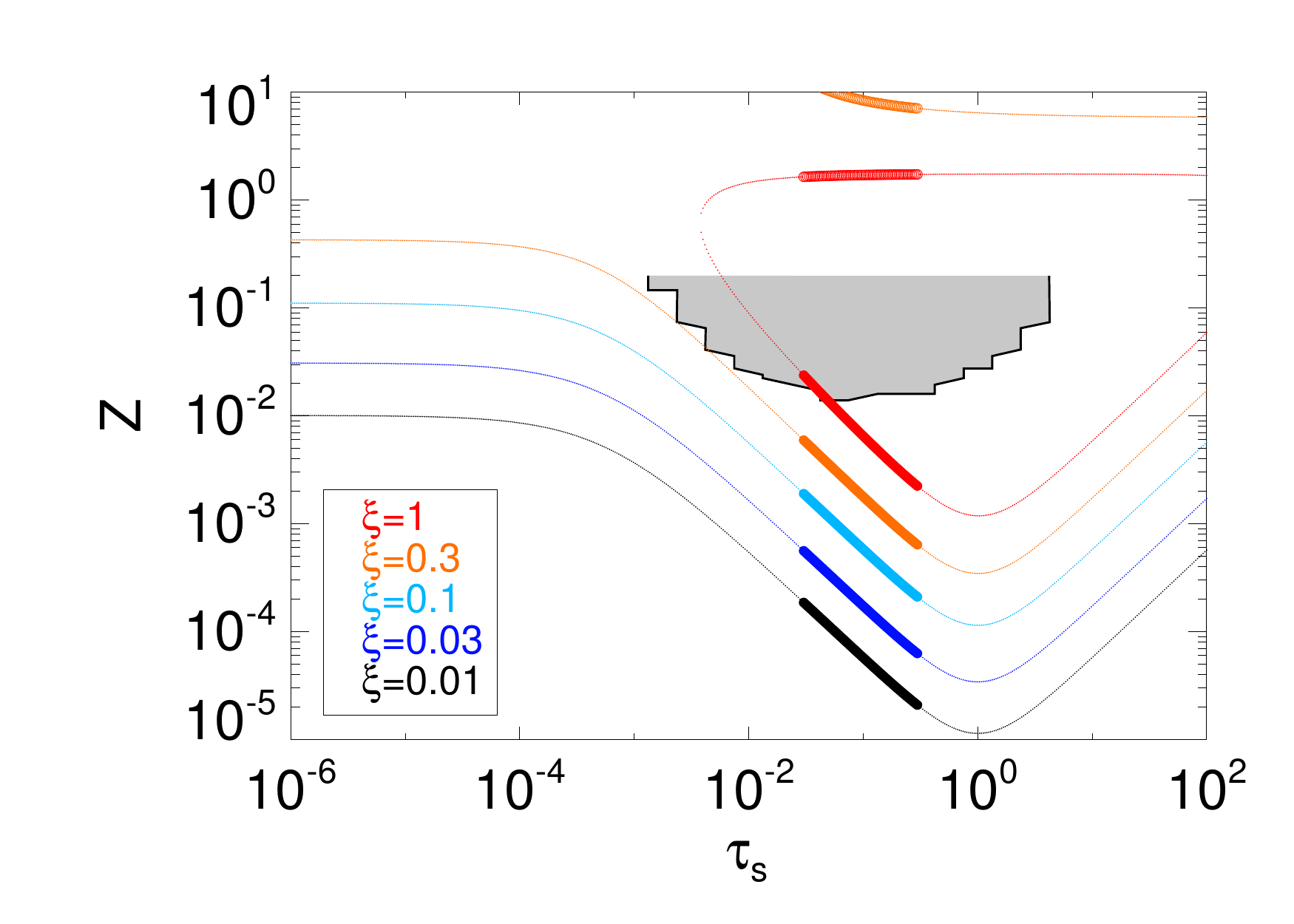}
\caption{Steady-state solutions for the solid-to-gas mixing ratio $Z$ as a function of the Stokes number of solid particles $\st$  for different values of the solid-to-dust-mass flux ratio $\xi$ (as labeled), assuming a value of the turbulent viscosity $\alpha=10^{-3}$. The values of $\st$ corresponding to expected pebble sizes are highlighted with larger symbols. The two solutions provided by Eq.~\eqref{eq:Z} are indicated by filled and open symbols, respectively. The gray area highlights the region in which planetesimals should form by a streaming instability \citep{Carrera15}.} \label{fig:Z_vs_taus}
\end{figure}

Figure~\ref{fig:Z_vs_taus} shows the solutions to Eq.~(\ref{eq:Z}) obtained for $Z$ as a function of $\st$ for different values of the pebble-to-gas-mass flux ratio $\xi$. 
 For pebbles, we expect $\st\sim 0.1$ \citep{Sato15,Ida16}, so that $\st\gg\alpha$ and hence $A/a_0 \simeq \beta^2 \gg 1$, yielding the approximate solution\begin{equation}
Z_{\rm peb}\sim \frac{\xi}{a_0}\alpha\frac{\st^2+1}{\st},
\label{eq:Zpeb}
\end{equation}
which fits the lower-$Z$ solutions in Fig.~\ref{fig:Z_vs_taus} for $\st \ga 10^{-2}$.
This approximate solution can be easily derived from 
Eq.~(\ref{eq:solid_to_gas}) with $\st\gg\alpha$ and $\Lambda \simeq 1$.
The dependence on $\alpha$ appears because, as shown by Eq.~\eqref{eq:Sigma}, $\sigg \propto 1/\alpha$ and 
$\sigp$ is independent of $\alpha$ for $\st \sim 0.1$.
The fast drift of pebbles is responsible for the small $Z_{\rm peb}$.

We expect planetesimal formation to occur either by GI in the dust disk 
when $\rho_{\rm p} \ga \rho_{\rm R}$, where $\rho_{\rm R} \sim M_*/r^3$ is the Roche density, independently of the dust size,
 or by SI with $Z$ as low as $\sim 0.02$ but a limited range of $\st$ values \citep[][also see Fig.~\ref{fig:Z_vs_taus}]{DD14, Carrera15}. 
In general, the condition for GI, $\rho_{\rm p} \ga \rho_{\rm R} $, is difficult to reach because of the vertical shear that
is due to KH instabilities and thus requires very high values of $Z$. 

We see in Fig.~\ref{fig:Z_vs_taus}  that the formation of pebbles by SI is possible but requires high values of $\xi$. 
The criterion for planetesimals to form directly from $\st\sim 0.1$ pebbles is 
$Z_{\rm pebble} \wig{>} 0.02$. From Eq.~(\ref{eq:Zpeb}), this implies
\begin{equation}
\xi\wig{>} \xi_{\rm crit,SI} \equiv 3.5 /\alpha_3,
\end{equation}
where $\alpha_3 = \alpha/10^{-3}$.
We show below that this condition, which requires the mass flux of pebbles to be equivalent to the mass flux of gas, is difficult to reach \citep[see also][]{Krijt16}. One possibility is to advocate high values of $\alpha$ \citep[see][]{Armitage+2016}, but this is generally not favored by the latest magnetohydrodynamical simulations of protoplanetary disks \citep[e.g.,][]{Bai2015}. Other possibilities exist that require local perturbations in the disk to modify the pressure gradient term \citep[e.g.,][and references therein]{JohansenPP6} or favorable conditions in terms of fragmentation threshold velocity and disk properties \citep{Laibe14,Drazkowska16}. 
We show below that the conditions necessary to form planetesimals can be reached at lower $\xi$ values and for small $\alpha$-disks next to a snow line. Before we examine this possibility, it is worth noting that for small particles (with $\st\wig{<}10^{-3}$), no solution is found for $\xi\ga 1$, meaning that  no steady-state exists: If there were a way to have small particles drift in at a very high rate or to deplete disk gas preferentially, the particles would pile up and accumulate, eventually forming planetesimals by direct gravitational instability.  

\section{Solid-to-gas density ratio inside of the snow line}

When the inward-drifting pebbles cross the snow line, they progressively sublimate until only small refractory (silicate dust) seeds remain \citep[e.g.,][]{saito_sirono11,  Morbidelli+2016}.  Observations of the interstellar medium and of interplanetary dust particles indicate that these dust seeds should be of submicron size, corresponding to $\st\sim 10^{-7}-10^{-5}$. Even for the much larger millimeter-sized chondrules, we expect $\st$ to be between $10^{-4}$ and $5\times 10^{-2}$ at most in a rarefied disk.
The presence of a snow line is thus a way of transforming a high-mass flux of fast-drifting pebbles into a flux of small, slow-drifting dust particles. 

Two additional factors need to be considered. First, the sublimation of the ice decreases the amount of solid material by a factor
$\zeta_0\sim 1/3$ corresponding to the ratio of the mass of dust (silicate components) to the total mass of condensates (dust+ice) \citep{Lodders2003}.  By assuming for simplicity that the pebbles instantaneously form small dust particles, the flux of material to inside of the snow line that is to be considered in Eq.~\eqref{eq:Z} is now therefore $\xi\rightarrow\zeta_0 \xi_{\rm peb}$, where $\xi_{\rm peb}$ corresponds to the ratio of the non-sublimated pebble mass flux to the gas mass flux.

Second, we expect dust grains to retain a memory of the vertical scale height of the pebbles. 
Their vertical mixing timescale can be estimated to be
\begin{equation}
t_{\rm mix} 
\sim (\hg/\ell)^2 \Omega_{\rm K}^{-1} 
\sim 160 \alpha_3^{-1} T_{\rm K} \sim 160 \alpha_3^{-1}  
(r/1{\,\rm au})^{3/2} {\rm yr},
\label{eq:t_mix}
\end{equation}
where $\ell \sim \sqrt{\alpha} \hg$ is the estimated vertical mean free path
and $T_{\rm K}$ is Kepler period.
Comparing a sublimation timescale with a migration timescale for pebbles,
we can derive the radial width for completion of the sublimation as
$\Delta r \sim 10^{-2}(R/10\,{\rm cm})^{1/2} r$ \citep[see also][]{Ciesla+Cuzzi2006}. 
With Eq.~(\ref{eq:Sigma}), the timescale for the pebble flux to establish $Z\ga 1$ in the sublimation region is estimated as
$t_Z \sim 2\pi r \Delta r \sigg/\dot{M}_{\rm peb} \sim (1/3\pi) (r/\hg)^2(\Delta r/r)\alpha^{-1}\xi_{\rm peb}^{-1} T_{\rm K}
\sim 10^3 (R/10\,{\rm cm})^{1/2} \alpha_3^{-1}\xi_{\rm peb}^{-1} T_{\rm K}$. Although $t_Z$ for $R=10$\,cm is 10-100 times longer than $t_{\rm mix}$,
the effective $R$ for sublimation would be much smaller and $t_Z$ would be much shorter for more realistic fluffy pebbles \citep[e.g.,][]{Kataoka+2013}.
We can thus assume that the dust seeds released by the sublimating pebbles have the same vertical thickness as the pebbles themselves. This is done in 
Eq.~\eqref{eq:Z}  by replacing $\beta$ by the value set by the pebble subdisk $\beta\rightarrow\beta_0\sim (1+\stpeb/\alpha)^{1/2}$. 

\begin{figure}[htb]
\includegraphics[width=\hsize,angle=0]{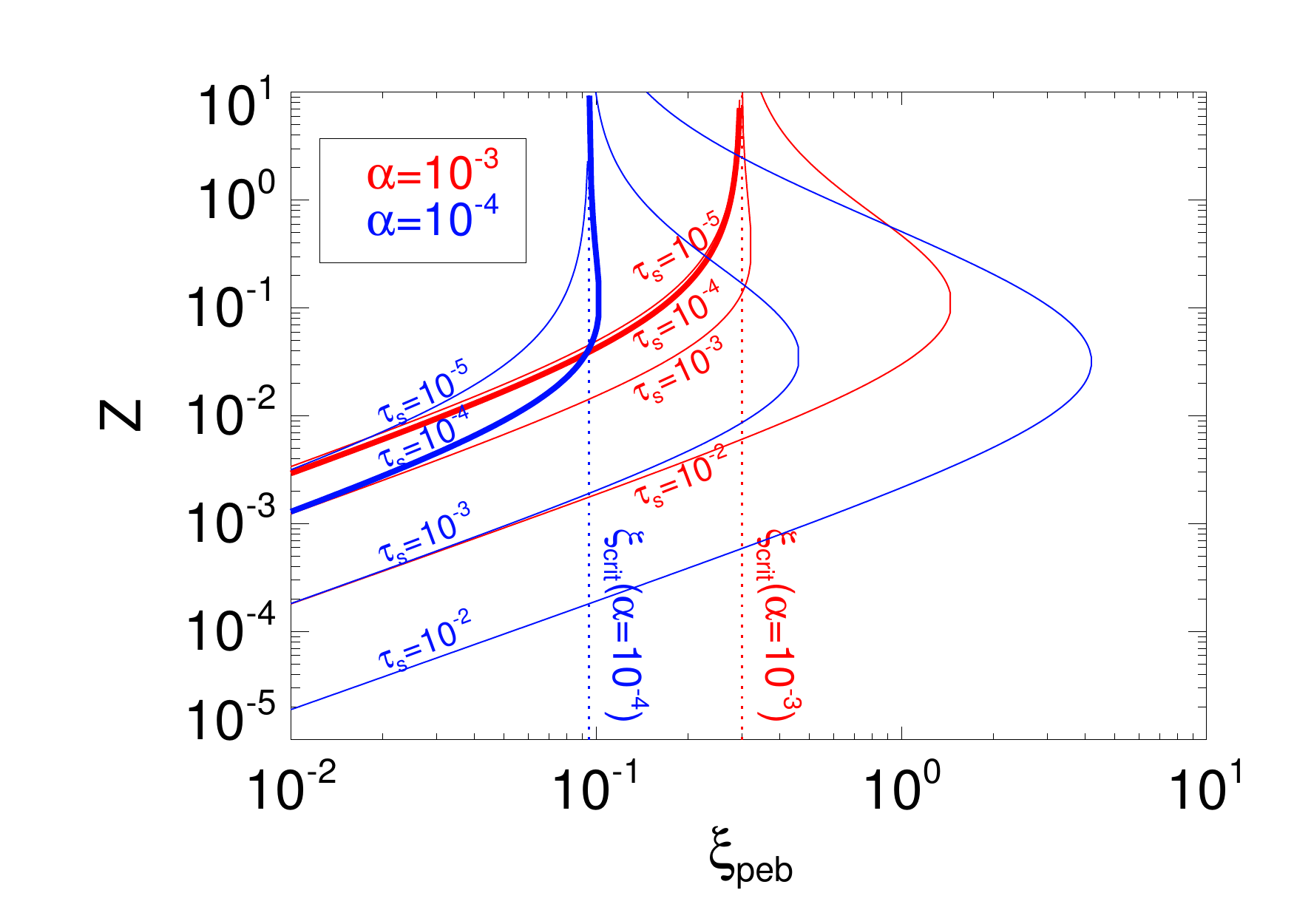}
\caption{Steady-state solutions for the solid-to-gas mixing ratio $Z$
  as a function of the solid-to-gas-mass flux ratio $\xi$ for
  different values of the Stokes number of solid particles $\st$ (from
  $10^{-5}$ to $10^{-2}$, as labeled), assuming two values of the turbulent viscosity $\alpha=10^{-4}$ (in blue) and $\alpha=10^{-3}$ (in red). 
  Equation~(\ref{eq:Z}) is numerically solved.
  In contrast to Fig.~\ref{fig:Z_vs_taus}, we now consider that initially icy pebbles with $\stpeb\sim 0.1$ and containing a mass fraction $\zeta_0=1/3$ in dust sublimate inside of the snow line. The thicker lines corresponds to the preferred value for the dust particles, $\st=10^{-4}$.}\label{fig:Z_vs_xi}
\end{figure}

As shown in Fig.~\ref{fig:Z_vs_xi}, the situation is now much more favorable toward the development of GI of the layer of sublimated pebbles. As pointed out by \citet{saito_sirono11} and \citet{Birnstiel12}, the release of small slow-drifting dust by fast-drifting pebbles naturally yields values of $Z$ that are much higher than for pebbles.
It first increases linearly with $\xi_{\rm peb}$.
However, when the pebble flux is high enough to yield $\rho_{\rm p}/\rho_{\rm g}\ga 1$, the dust migration velocity decreases as a result of the inertia of the dust, which further enhances $\rho_{\rm p}/\rho_{\rm g}$. The value of $Z$ then increases more than linearly with $\xi_{\rm peb}$ until the positive feedback becomes so strong that no steady-state solution can be found. At this point, we expect $\rho_{\rm p}$ to rapidly exceed $\rho_{\rm R}$ , and planetesimals are formed by GI. 

It is useful at this point to rederive Eq.~(\ref{eq:Z}) for these small grains, that is, in the limit of $A \ll 1$ and $\st \ll 1$:
\begin{equation}
Z \sim \frac{\zeta_0 \xi_{\rm peb}}{1-\beta \zeta_0 \xi_{\rm peb}}.
\label{eq:Z2}
\end{equation} 
For these small grains, the critical value $\xi_{\rm crit}$ above which no steady-state solution exists can be easily derived by calculating when the denominator of Eqs.~(\ref{eq:Z}) or~(\ref{eq:Z2}) becomes zero, that is, 
\begin{equation} 
\xi_{\rm crit} = \frac{1}{\beta \zeta_0} = \frac{1}{\zeta_0 (1+\tau_{\rm s,peb}/\alpha)^{1/2}}
\simeq \frac{1}{\zeta_0}\left(\frac{\alpha}{\tau_{\rm s,peb}}\right)^{1/2}.
\label{eq:xi_crit}
\end{equation}
The critical value $\xi_{\rm crit}$ is thus independent of the local pressure gradient. This is 
because the small dust grains ejected from sublimating pebbles are coupled to the disk gas motion and 
migrate with disk gas accretion and not by gas drag.

As shown by Fig.~\ref{fig:Z_vs_xi}, 
higher values of $\xi_{\rm crit}$ are possible for $\st \gg \alpha$, a possibility that we do not consider here because we expect seed grains to be such that $\st \la 10^{-5}$.
Typically, we thus obtain $\xi_{\rm crit}\simeq 0.3$ for $\alpha=10^{-3}$, $\zeta_0=1/3$ and $\stpeb=0.1$. The fact that $\xi_{\rm crit}$ scales with $\alpha^{1/2}$ is governed by the height of the dust and pebble subdisk. The formation of dust-rich planetesimals inside of the snow line is thus favored in weakly turbulent disks.

\section{Pebble flux and planetesimal formation}

The pebble mass flux is calculated by the mass in dust 
swept by the pebble formation front at $r \simeq r_{\rm peb}$ per unit time
\citep{LJ14b,Ida16},
\begin{equation}
\dot{M}_{\rm peb} {\displaystyle \simeq 2\pi r_{\rm peb} 
\times Z_0 \Sigma_g(r_{\rm peb}) \times \frac{dr_{\rm peb}}{dt}},
 \label{eq:M_F_est}
\end{equation}
where $Z_0$ is the solid-to-gas ratio in the pebble formation region.
The growth time of pebbles from $\mu$m sized dust grains is given by
\citep{Takeuchi+Lin05, Okuzumi12, Ida16}
\begin{equation}
t_{\rm grow} \sim 10 \, \times \, \frac{4}{\sqrt{3\pi}} \frac{1}{Z_0 \Omega}\sim 210 \; Z_{02}^{-1} M_{*0}^{-1/2}
\left(\frac{r_{\rm peb}}{1\,\au}\right)^{3/2}\,{\rm yr},
\label{eq:t_grow}
\end{equation}
where $Z_{02} = Z_0/0.01$. 
From this equation, the pebble formation front radius is given by
\begin{equation}
 r_{\rm peb} \simeq 290 Z_{02}^{2/3} M_{*0}^{1/3}t_6^{2/3}\,\au,
\end{equation}
where $t_6 = t/10^6{\rm yr}$.
Because ice needs to condense, $r_{\rm peb}\ga 1 \,\au$, implying
that pebbles may start forming when $t \ga 200\,{\rm yr}$.
Substituting this relation into Eq.~(\ref{eq:M_F_est}) with
$\dot{r}_{\rm peb}/r_{\rm peb} = (2/3t_{\rm grow})$, we obtain
\begin{equation}
\begin{array}{ll}
\xi_{\rm peb,pf} & 
{\displaystyle = \frac{\dot{M}_{\rm peb,pf}}{\dot{M}_*}
= \frac{\dot{M}_{\rm peb,pf}}{3 \pi \sigg \alpha \hg^2 \Omega} 
= \frac{\sqrt{3\pi}}{90}\frac{Z_0^2}{\alpha} \left(\frac{r_{\rm peb}}{\hg}\right)^2}  \\
 & {\displaystyle 
 \simeq 0.30 L_{*0}^{-2/7}M_{*0} \alpha_3^{-1}  Z_{02}^{34/21}
t_6^{-8/21}.}
\end{array}
  \label{eq:xi_pf}
,\end{equation}
where $L_{*0}=L_*/L_\odot$, $M_{*0}=L_*/M_\odot$,
$\alpha_3 = \alpha/10^{-3}$, 
 and we considered the irradiation-dominated regime with 
$T \simeq 120 L_{*0}^{2/7}M_{*0}^{-1/7} (r/1\,\au)^{-3/7}\; {\rm K}$,
for the pebble formation region in the outer disk. 
This corresponds to the disk aspect ratio,
\begin{equation}
\frac{\hg}{r} \simeq 
0.021 L_{*0}^{1/7}M_{*0}^{-4/7} \left(\frac{r}{1\,\au}\right)^{2/7}.
\label{eq:h_irr}
\end{equation}
(The expressions are slightly simplified compared to \citet{Ida16}.)
The pebble mass flux is given by $\xi_{\rm peb,pf} \dot{M}_{*}$, implying
\begin{equation}
\dot{M}_{\rm peb,pf} \simeq 0.90 \times 10^{-3} L_{*0}^{-2/7}M_{*0} \alpha_3^{-1}  Z_{02}^{5/3}\dot{M}_{*8}
t_6^{-1/3}  
\;\mearth/{\rm yr},
  \label{eq:M_F_est2}
\end{equation}
where $\dot{M}_{*8} = \dot{M}_*/(10^{-8} M_\odot/{\rm yr})$. This mass flux is higher than obtained by \citet{LJ14b} 
because they assumed a lower initial surface density in the gas disk ($\Sigma_{\rm g}=500\rm\,g/cm^2$ at 1\,au, not tied to the mass flux in the disk) and a $1/2$ coagulation probability.

When $r_{\rm peb}$ exceeds the disk size $r_{\rm out}$, 
that is, when $t \ga 2 \times 10^5 (r_{\rm out}/100\,\au)^{3/2}\ {\rm yr}$,
we would expect $\mdotf$ to decay more rapidly than $\dot{M}_*$ 
because most of the solid material has been made into pebbles and drifted in \citep{Sato15}. 
This may be inconsistent with the observational data that show that
mm or cm sized particles survive in the disks for several million years \citep{Brauer07}.
However, we expect
strong turbulence due to GI of the {\it \textup{gas disk}} (not GI of the {\it \textup{dust subdisk}}) 
to limit this fast spread of the pebble front, 
which could explain the presence of  mm or cm sized particles for relatively long times.

Disks with $\sigg$ given by Eq.~(\ref{eq:Sigma}) are gravitationally unstable in their outer parts
unless the disk is compact.
In the unstable parts, their surface density should evolve to become marginally unstable so that 
$1 \sim Q = c_s\Omega/\pi G \Sigma_{\rm g,GI}$, or equivalently
\begin{equation} 
\Sigma_{\rm g,GI} \simeq Q^{-1} \frac{h_{\rm g}}{r} \frac{M_*}{\pi r^2},
\label{eq:GI}
\end{equation}
where $Q$ is a factor on the order of unity. 
In these regions, the turbulence generated by
gravitational waves (assumed to lead to $\alpha_{\rm GI}\sim 0.1$) may
be high enough that collisions between icy grains result in fragmentation rather than coalescence. 
Typical collision velocities are estimated to be $v_{\rm col}\sim \sqrt{3 \alpha_{\rm GI} \st} c_s$ \citep{Sato15}. Assuming a sound speed $c_s \sim 270 (r/100 \,\au)^{-3/14}$\,m/s implies $v_{\rm col}\sim 50 (\alpha_{\rm GI}
/0.1)^{1/2}(\st/0.1)^{1/2} (r/100 \,\au)^{-3/14}$\,m/s. 
The threshold
velocity for fragmentation of icy particles
is predicted to be around $20-100$\,m/s \citep[e.g.,][]{Blum00, Zsom11,Wada11}.
We therefore assume that pebbles may form only in the stable
parts of the disks, that is, for $r<r_{\rm GI}$, implying that the location of the pebble formation
front is given by $\min(r_{\rm peb},r_{\rm GI})$. 

Since $\sigg$ given by Eq.~(\ref{eq:Sigma}) is equal to $\Sigma_{\rm g,GI}$ at $r = r_{\rm GI}$, 
we obtain $Q (r_{\rm GI}/h_{\rm g})^3 \simeq 3 \alpha (M_*/\dot{M}_*) \Omega$,
where $\alpha$ is the turbulence parameter for the inner regions.
With Eq.~(\ref{eq:h_irr}),
\begin{equation} 
r_{\rm GI} \simeq 160 L_{*0}^{2/3}M_{*0}^{-1/3}(\alpha_3/Q \dot{M}_{*8})^{14/9}\au.\label{eq:r_GI}
\end{equation}
As $\dot{M}_*$ decreases with time,
$r_{\rm GI}$ increases.
From $(dr_{\rm GI}/dt)/r_{\rm GI} \sim - (14/9)(d \dot{M}_*/dt)/\dot{M}_* $
and Eq.~(\ref{eq:GI}),
the pebble mass flux that is due to the outward spread of $r_{\rm GI}$ is
\begin{equation}
\begin{array}{ll}
\xi_{\rm peb,GI} & {\displaystyle  
= \frac{2\pi r_{\rm GI} \times Z_0 \Sigma_g \dot{r_{\rm GI}}  }{\dot{M}_*}
\simeq - \frac{14}{9} Z_0 \frac{h_{\rm g}}{r_{\rm GI}} \frac{M_*}{\dot{M}_*^2} \frac{d \dot{M}_*}{dt}. }   \end{array}
  \label{eq:xi_GI}
\end{equation}
We now use the relation between accretion rate and age suggested by the observations of young clusters, $\dot{M}_{*8} \sim t_6^{-3/2}$ \citep{Hartmann+1998}. This yields
\begin{equation} 
r_{\rm GI} \simeq 160 L_{*0}^{2/3}M_{*0}^{-1/3} (\alpha_3/Q)^{14/9} t_6^{7/3}\ \au
\end{equation}
and
\begin{equation}
\xi_{\rm peb,GI} \simeq 0.24 L_{*0}^{1/3}M_{*0}^{1/3} Z_{02}
 (\alpha_3/Q)^{4/9} t_6^{7/6}.
 \label{eq:xi_pebGI}
\end{equation}
Equivalently,
\begin{equation}
\dot{M}_{\rm peb,GI}
\simeq 0.79 \times 10^{-3} L_{*0}^{1/3}M_{*0}^{1/3}Z_{02}
 (\alpha_3/Q)^{14/9} t_6^{-1/3} \;\;\mearth/{\rm yr}.\end{equation}

\begin{figure}[htb]\
\includegraphics[width=75mm,angle=0]{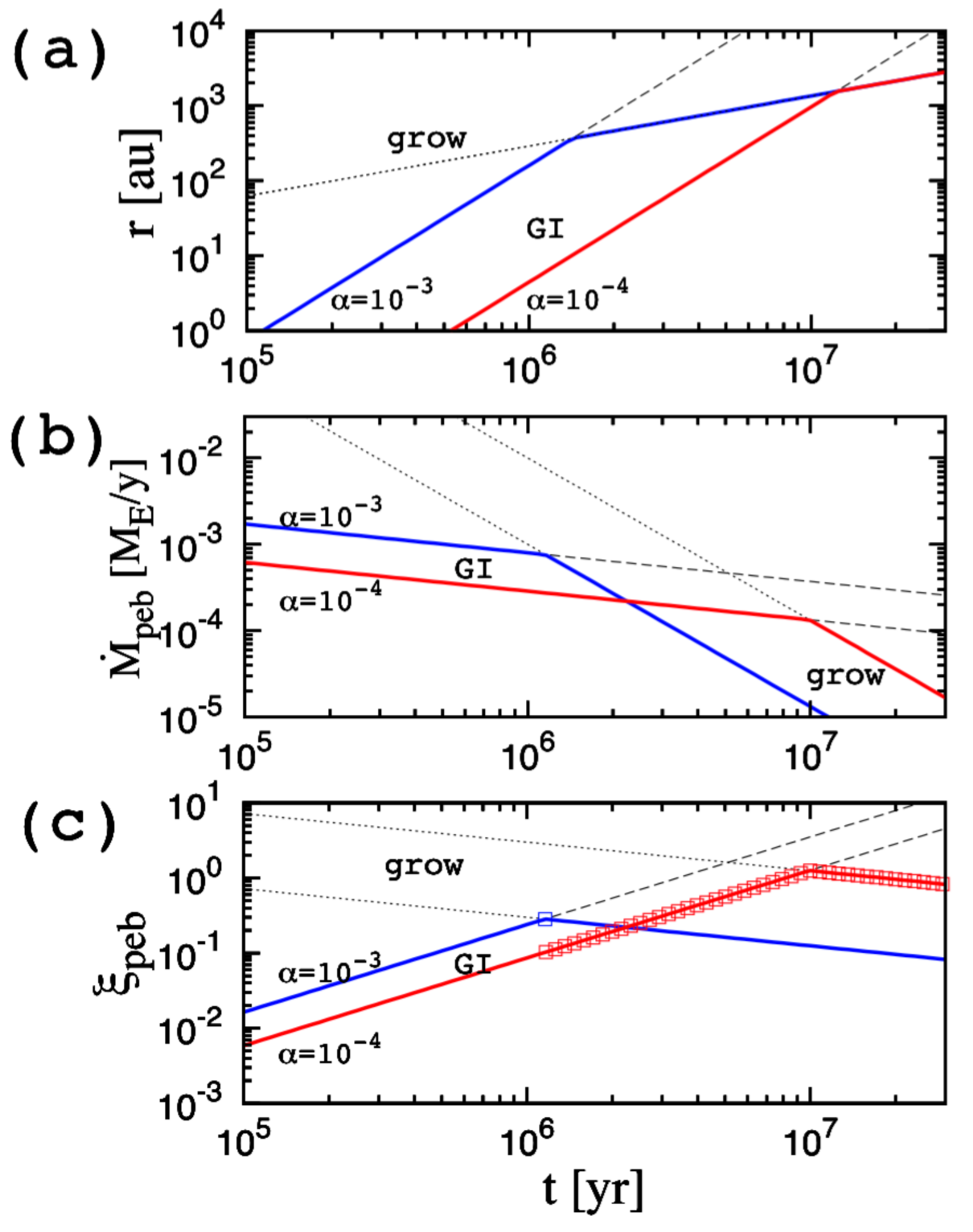}
\caption{Time evolution of 
(a) radius of the pebble formation front, (b) pebble accretion rate ($\mdotf$), and
(c) $\xi_{\rm peb} = \dot{M}_{\rm peb}/\dot{M}_{*}$ for two values of $\alpha$,
$10^{-4}$ (red) and $10^{-3}$ (blue). 
The lines labeled "grow" (dotted) and "GI" (dashed) represent 
the pebble growth and disk GI limits, respectively.
The thick solid lines express the actual values obtained by the minima of
the two limits.
Here we assumed $\stpeb = 0.1$ and $Z_0 = 0.01$.
In panel c, 
the small squares represent the points with $\xi_{\rm peb} > \xi_{\rm crit}$ , see Eq.~\eqref{eq:xi_crit}.
}
\label{fig:tevol}
\end{figure}

Figure~\ref{fig:tevol} shows the evolution of the pebble formation
front, the pebble mass flux $\dot{M}_{\rm per}$ , and $\xi_{\rm peb}$. 
The mass flux ratio $\xi_{\rm peb}$ reaches a maximum when
$r_{\rm GI} = r_{\rm peb}$. 
This maximum value is proportional to $\alpha^{-29/45}$,
whereas $\xi_{\rm crit}\propto \alpha^{1/2}$. 
For the
$\alpha=10^{-3}$ case, we see that $\xi_{\rm peb}$ becomes equal to
$\xi_{\rm crit}$ obtained from Eq.~\eqref{eq:xi_crit} only for a very
short time. The $\alpha=10^{-4}$ case, in contrast, leads to a
prolonged period in which planetesimals can form inside the snow line. Thus, direct formation of planetesimals is
possible in low-turbulence disks. Conversely, in high-turbulence
disks, the relatively low maximum values of $\xi_{\rm peb}$ obtained imply that other mechanisms have to be sought so that planetesimals
can form. This may involve disk photoevaporation, disk winds, or growth of pebbles by ice condensation.
 
When formed by GI, the mass of a clump is $M_{\rm clump}\sim \hp^3 \rho_{\rm R}$,
where $\rho_{\rm R}\sim M_*/r^3$ is the Roche density.
If the clump shrinks into a planetesimal with physical radius $R$ and
bulk density $\rho_s \,(\sim 1{\rm gcm}^{-3}$), then for the solar case,
\begin{equation}
R \sim \left(\frac{\hp}{r}\right)\left(\frac{M_*}{\rho_s}\right)^{1/3} 
\sim 10^3 \left(\frac{\hg/r}{0.01}\right)
\alpha_3^{1/2}
\left(\frac{\stpeb}{0.1}\right)^{-1/2}\ {\rm km}.
\end{equation} 
Although a more detailed analysis would be required, it appears that
the planetesimals formed through this mechanism are as large as those
formed by SI \citep[e.g.,][]{JohansenPP6}.  


\section{Discussion and conclusion}

With a simple model of pebble growth, drift, and sublimation at the
snow line, we have examined the conditions for
the formation of planetesimals  in protoplanetary disks. 

We first showed that forming planetesimals from streaming instability in the flow of icy pebbles requires both a high level of
turbulence ($\alpha\ga 0.01$) and an unrealistically high pebble
flux. We note that for these high turbulence levels, water vapor
diffusion can decrease the requirement on the pebble mass flux
\citep[e.g.,][]{Ros_Johansen,Armitage+2016}.

By including the often neglected mass-loading factor in the equations
for the drift of solids, we have shown that the pile-up of solids inside of the snow line 
leads to the formation of dust-rich planetesimals
directly by gravitational instability in the dust subdisk. This
instability exists for relatively high values of $\xi$, the pebble-to-gas-mass
flux ratio, and for relatively low values of $\alpha$. 

With a simple model of the formation of pebbles, we have demonstrated
that super-critical values of the pebble-to-gas mass flux $\xi\ge\xi_{\rm crit}$ are reached in disks with $\alpha\la 10^{-3}$. This condition may be reached more easily, that is, for higher values of $\alpha$ or lower values of $\xi$, by taking into account disk gas depletion mechanisms other than viscous disk accretion such as photoevaporation or disk winds. 
The planetesimals that are formed in this way are expected
to be large, probably larger than 100\,km in radius. 

During that time, the H$_2$O snow line could move from several $\au$ to inside of $1\;\au$ \citep[e.g.,][]{Oka+2011}. 
These planetesimals are expected to be composed of a high
fraction of dust (silicates), which may explain why the rock-to-ice
fractions inferred in minor planets and moons in the outer 
solar system \citep[e.g.,][]{Schubert+2010} or the dust-to-ice ratio in
comets \citep[e.g.,][]{Rotundi+2015, Lorek+2016} are often significantly
higher than the expected $1/2$ to $1/3$ value obtained from purely solar
composition \citep[e.g.,][]{Lodders2003}. 

Our model requires a fast breakup of pebbles, however, so that the dust
particles are released over a small annulus. Given that we expect
these pebbles to be porous \citep[e.g.,][]{Kataoka+2013}, this should
be verified. We note that the possibility that disks have flow in the
midplane that are directed outward \citep{Takeuchi+Lin05} or are stochastic
\citep{Suzuki+2014} will favor the mechanism that we propose.

Finally, this process may apply to other sublimation lines
\citep[e.g.,][]{Drozdovskaya+2016} if these 
lead to the breakup of fluffy pebbles into much smaller
grains. Indeed, sintering has been shown to have this effect and thus
might explain the rings that were recently observed in young disks
\citep{Okuzumi+2016}. The pile-up and planetesimal formation
  mechanism that we propose may thus naturally explain the formation
  of rings of planetesimals in low-turbulence disks.

\begin{acknowledgements}
We thank Chris Ormel, Anders Johansen, Satoshi Okuzumi, and an anonymous referee for helpful comments.
S. I. thanks for the hospitality he experienced during his visit to the Observatoire de la C\^ote d'Azur, which was made possible thanks to support from {\em OCA BQR}. S. I. is also supported by MEXT Kakenhi grant 15H02065.
We acknowledge support by the French ANR, project number ANR-13-13-BS05- 0003-01 projet MOJO (Modeling the Origin of JOvian planets)
\end{acknowledgements}

\bibliography{si} 

\end{document}